\documentclass[journal=ancac3,manuscript=article]{achemso}
\usepackage{chemformula} % Formula subscripts using \ch{}
\usepackage[T1]{fontenc} % Use modern font encodings
\usepackage{mhchem}
\usepackage{hyperref}
\usepackage{bm}
\usepackage{braket}
\usepackage{physics}
\usepackage{multirow}
\usepackage{framed}
\SectionNumbersOn
\newcommand{\kB}{\mathrm{k}_\mathrm{B}}
\newcommand{\kQTST}{k_\mathrm{QTST}}

\newcommand{\kRPMD}{k_\mathrm{RPMD}}
\newcommand{\kcl}{k_\mathrm{cl}}

\author{Somnath Bhowmick}
\affiliation{Computation-based Science and Technology Research Center, The Cyprus Institute, 20 Konstantinou Kavafi Street, Nicosia 2121, Cyprus.}
\email{s.bhowmick@cyi.ac.cy}
\author{Marta I. Hern{\'a}ndez}
\affiliation{Instituto de Física Fundamental, Consejo Superior de Investigaciones Científicas (IFF-CSIC), Serrano 123, 28006 Madrid, Spain}
\author{Jos\'{e} Campos-Mart\'{i}nez}
\affiliation{Instituto de Física Fundamental, Consejo Superior de Investigaciones Científicas (IFF-CSIC), Serrano 123, 28006 Madrid, Spain}
\author{Yury V. Suleimanov}
\affiliation{Computation-based Science and Technology Research Center, The Cyprus Institute, 20 Konstantinou Kavafi Street, Nicosia 2121, Cyprus.}
\alsoaffiliation{Department of Chemical Engineering, Massachusetts Institute of Technology, 77 Massachusetts Ave., Cambridge, Massachusetts 02139, United States}
\email{y.suleymanov@cyi.ac.cy}
\phone{+357 22208721}
\fax{+357 22208625}

\title[]
  {Isotopic Separation of Helium through Nanoporous Graphene Membranes: A Ring Polymer Molecular Dynamics Study}

\begin{document}

\begin{abstract}
Microscopic-level understanding of the separation mechanism for two-dimensional (2D) membranes is an active area of research due to potential implications of this class of membranes for various technological processes. Helium (He) purification from the natural resources is of particular interest due to the shortfall in its production. In this work, we applied the ring polymer molecular dynamics (RPMD) method to graphdiyne (Gr2) and graphtriyne (Gr3) 2D membranes having variable pore sizes for the separation of He isotopes. We found that the transmission rate through Gr3 is many orders of magnitude greater than Gr2. The selectivity of either isotope at low temperatures is a consequence of a delicate balance between the zero-point energy effect and tunneling of $^4$He and $^3$He. RPMD provides an efficient approach for studying the separation of He isotopes, taking into account quantum effects of light nuclei motions at low temperatures, which classical methods fail to capture.
\end{abstract}

\section{Introduction}
The use of two-dimensional (2D) materials is an open field for many technological applications as well as for basic science\cite{karin-2dmat:20,research-graph2yne:18}. Graphene, one of the pioneer materials that started this very active area of research, acts as an almost impermeable sheet for most atomic and molecular species due to its high electron density surrounding the aromatic rings.\cite{McEuen_NL:2008,Berry_Carbon:2013} Some years ago, experiments by Lozada-Hidalgo \emph {et al.}\cite{Lozada-hidalgo-1st-proton-permea:14,Hidalgo_Science:2016} and recently by Creager \emph{et al.}\cite{Creager_EA:2019} have shown that small charged particles such as the proton and its isotopes can penetrate through a monolayer of pristine graphene. There is no complete explanation despite many theoretical calculations, \cite{Mazzuca_JCP:2018,Liu_PCCP:2013,Bartolomei_Carbon:2019} although some kind of chemical interaction seems to be responsible for this process.  Even more recently, it has also be found that hydrogen can permeate pristine graphene \cite{LozadaHidalgo-hydrogen-permea:19}.  However, the penetration energy barrier for other neutral atoms is significantly higher and therefore supports the widely recognized notion of graphene impermeability.\cite{Tsetseris_Carbon:2014} Introducing defects and moderate annealing into the membrane, and nanopores of different sizes, may enable graphene and other 2D layers to act as atomic and molecular sieves. \cite{Tsetseris_Carbon:2014,Zhao_JPCB:2006,Koeing_NN:2012,Jiang_NL:2009,Hauser_JPCL:2012,Joshi_Science:2014,Swathi_JPCB:2018} The use of 2D materials as a filter at the molecular level is one of the most interesting applications.\cite{Koeing_NN:2012} Besides using 2D membranes with fabricated nanopores, some compounds contain nanopores in their structures at different positions and sizes; therefore, they are more naturally suited for filtering at the molecular level.

An interesting class of 2D nanoporous material is graphyne,\cite{Yeo_AM:2019,graphyne-family:19} which is composed of $sp-sp^2$ hybridized C atoms and can be considered as a graphene derivative. In graphyne, one-third of C$-$C bonds have been replaced by mono- and poly-acetylenic units (\ce{-C#C- })$_N$. Baughman \emph{et al.},\cite{Baughman_JCP:1987} theoretically proposed the first stable structures of graphyne, which was two decades later synthesized on a copper substrate.\cite{Li_CC:2010,Zhou_JACS:2015} $N$ defines the number of acetylenic linkages and, consequently, the size of the uniformly distributed and repeating sub-nanometer triangular pores. They are termed as graph-$N$-yne membranes, such as graphdiyne (Gr2), graphtriyne (Gr3), \emph{etc}., for $N$ = 2 and 3, respectively.\cite{Bartolomei_JPCL:2014} The chemical and mechanical properties of these graphyne membranes have many useful features, such as they are chemically inert and stable at ambient temperatures,\cite{Cranford_Carbon:20111,Ivanovskii_PSSC:2011,Buehler_Nanoscale:2012} and flexible enough to withstand deformations induced by high pressures.\cite{Yang_CMS:2012,Buehler_Nanoscale:2013} The above-mentioned properties of graphynes, coupled with their unique geometrical structure, make them an excellent candidate to be utilized in gas separation and water filtration technologies. Indeed, one can find many reports on the purification of gases such as H$_2$,\cite{Jiao_CC:2011} N$_2$,\cite{Zhao_ASS:2017} and O$_2$\cite{Meng_AMI:2016} from a mixture of gases and desalination and filtration of water\cite{Bartolomei_JPCL:2014,Buehler_Nanoscale:2013,Zhu_SR:2013} in the literature. 

There is a growing worldwide demand for helium purification from its natural resources due to the shortfall in its production.\cite{Nuttall_Nature:2012} It has numerous industrial and scientific applications such as superconducting magnets, space rockets, arc welding, \emph{etc}. In particular, the lighter isotope, $^3$He, also plays a pivotal role in fundamental research, for example, in neutron-scattering centers, ultracold physics and chemistry, \emph{etc}.\cite{Bartolomei_JPCC:2014} The relative abundance of $^3$He is low ($\approx$1.34$\times$10$^{-6}\;\;$\%)\cite{atomic_weight} in comparison to its heavier isotope, and its extraction from natural gas is usually done by means of expensive cryogenic distillation and pressure-swing adsorption methods.\cite{Das_CC:2008} An alternate and more energy-efficient process is to use the 2D porous membranes in isotopic gas separation since they usually do not involve costly liquefaction of the gases.\cite{Bernardo_IECR:2009} Many theoretical works have been reported in the last decade on the separation of He isotopes using graphene derivatives, such as: polyphenylene (2D-PP),\cite{Blankenburg_Small:2010,Schrier_JPCL:2010,Schrier_CPL:2012,Schrier_JPCC:2013,Bartolomei_JPCC:2014} functionalized graphene pores,\cite{Hauser_JPCL:2012,Schrier_JPCC:2012,Bhatia_JPCC:2015,Zhu_CMS:2019} nanoporous multilayers,\cite{Schrier_JPCA:2014,Bhatia_JPCC:2015} Gr2,\cite{Bartolomei_JPCC:2014,Hernandez_JPCA:2015,Hernandez_JPCC:2017} Gr3,\cite{Marta:2021} holey graphene \cite{Hernandez_JPCC:2017}, and graphenylene membranes, \cite{Zhao_PCCP:2017} \emph{etc}.

The light element He within the vicinity of subnanometer pores is an obvious environment for observing the important role of the quantum mechanical effects such as zero-point energy (ZPE) and tunneling effects. A wise approach for an effective isotopic separation maybe is to exploit these quantum properties that could run in the opposite directions. The heavier isotope with smaller ZPE will diffuse faster, while quantum mechanical (QM) tunneling favours the lighter species. Previously, some of us, using quantum three-dimensional wave packet calculations have shown that the $^4$He/$^3$He selectivity increases with decreasing temperature for Gr2 and holey graphene membranes.\cite{Hernandez_JPCA:2015,Hernandez_JPCC:2017} It has also been reported that the effect of ZPE is more dominant than tunneling at low temperatures (20$-$40 K).\cite{Hernandez_JPCA:2015} On the contrary, within a low but acceptable gas flux and at low temperatures (10$-$30 K), separation on various functionalized membranes indicates increased selectivity for the lighter isotope due to QM tunneling.\cite{Schrier_JPCC:2012,Hauser_JPCL:2012,Bhatia_JPCC:2015,Zhao_PCCP:2017}

The ring polymer molecular dynamics (RPMD) method,\cite{Craig_JCP_2:2005,Craig_JCP_1:2005,Habershon_ARPC:2013,Suleimanov_JPCA:2016} based on the imaginary-time path integral formalism, is an efficient approach that can accurately and reliably  describe the ZPE\cite{Tudela_JPCL:2012} and deep quantum tunneling effects\cite{Craig_JCP_2:2005,Richardson_JCP:2009,Tudela_JPCL:2014}. RPMD method is essentially a classical molecular dynamics method in an extended ring polymer phase space. It can provide reliable estimates of thermal rate coefficients since the RPMD partition function rigorously converges to the QM partition function\cite{Craig_JCP:2004}, the long-time limit of the ring polymer flux-side correlation functions is independent of the choice of the dividing surface that separates reactants from products, while its short-time limit is related to various quantum transition state theories\cite{Suleimanov_JPCA:2016}. RPMD was introduced in an \emph{ad hoc} manner by Craig and Manolopoulos to study the dynamics of the condensed phase processes,\cite{Craig_JCP_2:2005,Craig_JCP_1:2005} owing to its simplicity and efficiency (scales favorably with the dimensionality of the system). Examples of its successful application include diffusion in and inelastic neutron scattering from liquid para hydrogen,\cite{Miller_JCP_1:2005,Craig_CP:2006} the translational and orientational diffusion in liquid water,\cite{Miller_JCP_2:2005} proton transfer in water,\cite{Mazzuca_JPCA:2017} diffusion of H and $\mu$ atoms in liquid water, hexagonal ice,\cite{Markland_JCP:2008} and on Ni surface,\cite{Suleimanov_JPCC:2012} electron transfer\cite{Menzeleev_JCP:2011} and proton-coupled electron transfer,\cite{Kretchmer_JCP:2013} enzyme catalysis,\cite{Boekelheide_PNAS:2011} \emph{etc}.  However, the RPMD method is not restricted to the condensed phases and also found wide application in calculating rate coefficients for the gas-phase bimolecular reactions\cite{Guevara_JCP:2009,Suleimanov_JCP:2011,Suleimanov_CPC:2013}  as explored by Suleimanov and co-workers, see for instance, a review by one of us\cite{Suleimanov_JPCA:2016} and a recent paper\cite{Bhowmick_PCCP:2018} and references  mentioned therein.  In the permeation of H$^+$ and D$^+$ on a pristine graphene, the RPMD method also has been used,\cite{Mazzuca_JCP:2018} and the authors point out, again, the importance of considering both ZPE and tunneling effects for isotopic separation.

Coupled with the major scientific and industrial appeal, in this paper, we examine the feasibility of separation of He isotopes using 2D Gr2 and Gr3 membranes at low temperatures (20$-$250 K) using the RPMD method. A significant part of the success of the RPMD method is attributed to the fact that it gives the exact quantum-mechanical rate coefficient for the transmission through a parabolic barrier,\cite{Craig_JCP_2:2005} which is advantageous for the present investigation. The primary objective of this study is to provide a reliable estimate of the selectivity and to show that RPMD is a necessary alternative to study these processes where quantum effects are expected to be very important. To the best of our knowledge, the RPMD method has not been applied for the He isotope separation previously and therefore presents an excellent opportunity to test the accuracy of this method.

The paper is organized as follows: in the next section (\autoref{sec:method}), we provide the details of the RPMD approach along with the PES used in the present study. The results of RPMD rate coefficients and selectivity have been compared with earlier studies\cite{Hernandez_JPCA:2015,Hernandez_JPCC:2017,Marta:2021} in \autoref{sec:result}. Concluding remarks are provided in the last section (\autoref{sec:conclusion}).

\section{Computational details}\label{sec:method}

\subsection{The simulation setup and potential energies}

We investigate the rate of transmission of both $^3$He and $^4$He through 2D graphdiyne (Gr2) and graphtriyne (Gr3) membranes using the RPMD method. The unit cell of Gr2 and Gr3 has the dimensions (in $x-$ and $y-$ coordinates) of (16.37 \AA, 9.45 \AA) and  (20.82  \AA, 12.02  \AA), respectively. A comprehensive study on the geometry of the membranes can be found in Ref.~\citenum{Bartolomei_JPCC:2014}. All molecular dynamics simulations have been performed on graphyne membranes containing using same periodic unit as in Ref.~\citenum{Hernandez_JPCC:2017}. 

The interaction potential between He$-$graphyne membranes is obtained as an additive improved Lennard-Jones (ILJ) He$-$C pair potentials. The optimized values of the parameters of the ILJ potentials have been determined from ``coupled'' supermolecular second-order M{\o}ller-Plesset perturbation theory (MP2C) theory\cite{Pitonak_JCTC:2010} using aug-cc-pVTZ and aug-cc-pV5Z basis set for C- and He- atoms respectively. These IJL potentials correspond to the same potentials as employed in previous studies of He transmission through graphyne membranes,\cite{Bartolomei_JPCC:2014,Hernandez_JPCA:2015,Hernandez_JPCC:2017} and a detailed description can be found elsewhere.\cite{Bartolomei_JPCC:2014,Hernandez_JPCA:2015}

\begin{figure}[!h]
   \centering
   \includegraphics[width=0.70\textwidth]{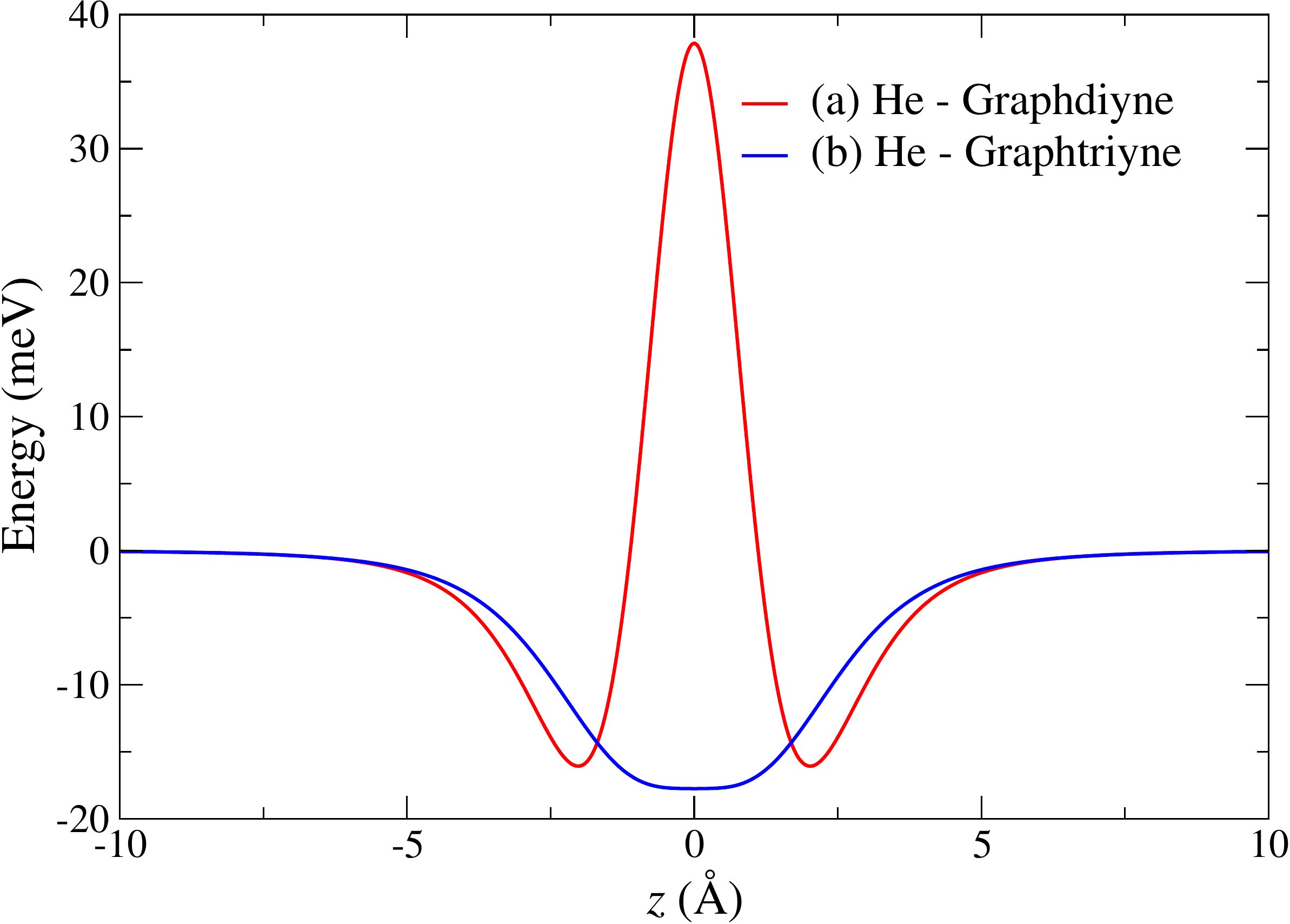} 
   \caption{Variation of the potential energies (in meV) for (a) He atom$-$graphdiyne and (b) He atom$-$graphtriyne interaction along the minimum energy transmission pathway (MEP) (in \AA). The interaction energies are calculated by using improved Lennard-Jones (ILJ) potentials. The MEP corresponds to He atom perpendicularly approaching the geometric center of graphdiyne and graphtriyne pores with reaction coordinates (0,0,$z$).}
\label{fig:PEC}
\end{figure}

\begin{figure}[!h]
   \centering
   \includegraphics[width=0.80\textwidth]{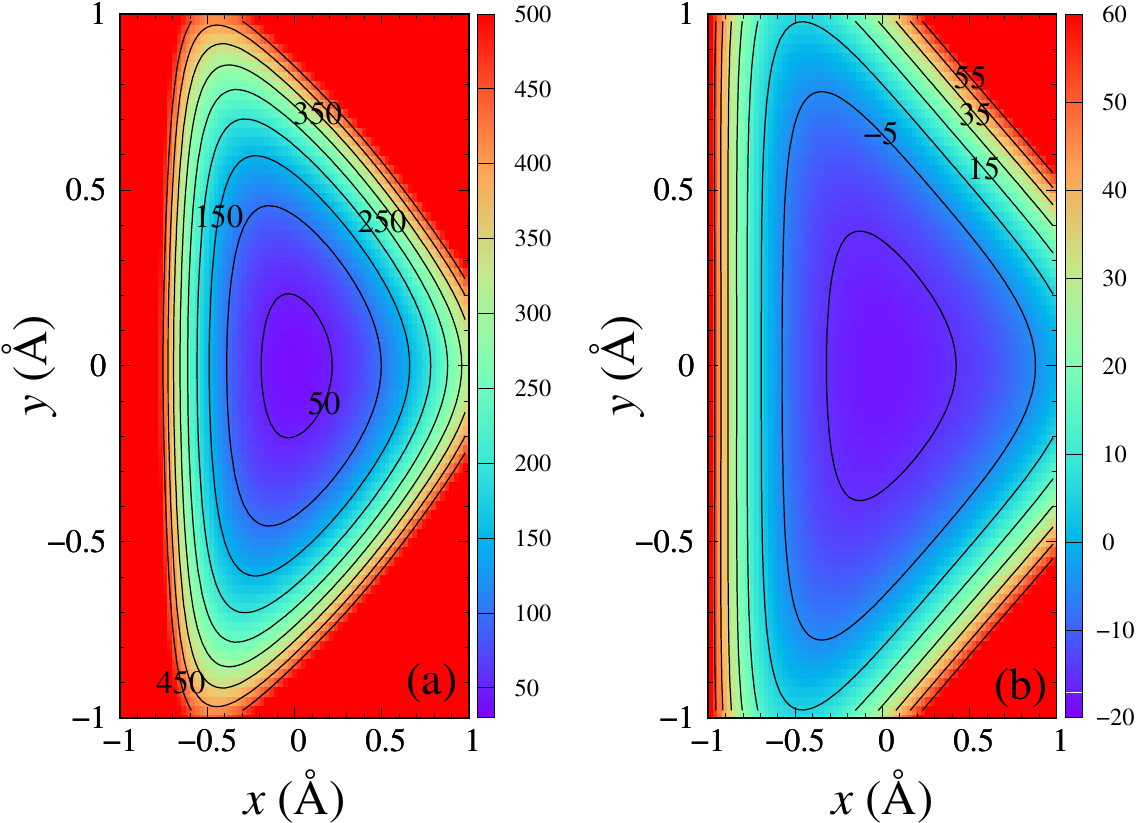} 
   \caption{Improved Lennard-Jones (ILJ) interaction potentials (in meV) for the displacement of He atom along $x$ and $y$ directions from the center of the pore (0,0,0) on (a) graphdiyne and (b) graphtriyne membranes. The $z$ coordinate of He atom was held at the origin. The energy contour step value is 50 meV and 10 meV for (a) and (b), respectively.}
\label{fig:IPD}
\end{figure}

The potential energy curves of He along the transmission pathway ($z-$ coordinates) to both membranes are illustrated in \autoref{fig:PEC}, while the contour plots in \autoref{fig:IPD}(a) and (b) highlight the in-pore displacements of He along $x-$ and $y-$ coordinates. It is quite evident from both these plots that the minimum energy path (MEP) for an effective He transmission may correspond to a straight line perpendicular to the center of the pore. For the transmission through the Gr2 membrane, the He atom lying at the center of a pore is the saddle point. The maximum potential barrier height for the MEP is 37.85 meV, which is similar to the previous studies.\cite{Bartolomei_JPCC:2014,Hernandez_JPCA:2015,Hernandez_JPCC:2017} In contrast, MEP for the He $-$ Gr3 PEC is devoid of any potential barrier; instead, a large well (of depth 17.74 meV) appears within the vicinity of the membrane.\cite{Marta:2021} These results suggest that He penetration of larger Gr3 pores should be much easier than that of smaller Gr2 pores, analogous to the report on water permeation by Bartolomei \emph{et al.}\cite{Bartolomei_JPCL:2014} Furthermore, along this MEP and He atom-membrane perpendicular distance of around 6.0 \AA, the interaction potential reaches a plateau and therefore can be characterized as an asymptotic reactant site. It is interesting to note that the interaction potential steeply rises to a  high value for any movement along the in-pore $x-$ and $y-$ degrees of freedom [see \autoref{fig:IPD}(a) and (b)].

\subsection{Ring polymer molecular dynamics (RPMD) method}\label{sec:rpmd}

The classical Hamiltonian for the system composed of $^4$He or $^3$He atom under the influence of external potential arising from $M$ carbon atoms of fixed Gr2 or Gr3 membrane can be written in the atomic unit as:

\begin{equation}
\hat{H}= \frac{|\hat{\bm{p}}|^2}{2m} + V(\hat{\bm{r}}_{\text{He}},\hat{\bm{r}}_{\text{C}(1)},...,\hat{\bm{r}}_{\text{C}(M)}),
\end{equation}

\noindent where, $\hat{\bm{p}}$ and $\hat{\bm{r}}_{\text{He}}$ (or $\hat{\bm{r}}_{\text{C}(i)}$) are the momentum and position vectors of the $i$th atom of mass $m_i$. In the RPMD method, the He atom is treated as $n$ classical replicas of the original particle, each connected with its nearest neighbor by harmonic springs. The modified ring polymer Hamiltonian has the form: 

\begin{equation}
H_n(\bm{p},\bm{r}) = H_n^0(\bm{p},\bm{r}) + V(\bm{r}_{\text{C(1)}},...,\bm{r}_{\text{C(M)}}) +  \sum_{j=1}^{n}V(\bm{r}_{\text{He}}^{(j)}),
\end{equation}

\noindent where,
\begin{equation}
H_n^0(\bm{p},\bm{r}) = \sum_{j=1}^{n}\left(\frac{|{\bm{p}}^{(j)}|^2}{2m} + \frac{1}{2}m\omega_n^2|\bm{r}^{(j)}-\bm{r}^{(j-1)}|^2\right).
\end{equation}

\noindent $n$ is the number of classical beads representing quantum He atom which are connected by a harmonic potential with force constant $\omega_n$ (=$\beta_n\hbar$). $\beta_n \equiv \beta/n$ is the reciprocal temperature of the system, $\beta = 1/\kB T$. $\kB$ is the Boltzmann constant, $T$ is the system temperature, and $\hbar$ is the Dirac constant ($\hbar$=1 in atomic units). $\bm{p}^{(j)}$
and $\bm{r}^{(j)}$ are the momentum and position vectors of the $j$th bead in the ring polymer necklace of He atom, respectively.  The ring polymer trajectory now evolves in $f=3n$ total degrees of freedom (in atomic Cartesian coordinates and including translation and rotational degrees of the freedom of the entire system which is a convenient method to propagate RPMD trajectories\cite{Suleimanov_JCP:2011}). 

We perform the RPMD simulations at temperatures $T$ below 250 K because, previously, it was observed that the maximum selectivity was obtained at low temperatures.\cite{Hernandez_JPCC:2017,Hernandez_JPCA:2015} We choose seven different $T$ = 20, 30, 50, 100, 150, 200, and 250 K, to study the variation of selectivities with $T$ and to find an optimal temperature range for isotopic separation.  Furthermore, the number of beads is a measure of the resolution of the path integral  calculations, \emph{i.e.}, RPMD calculations scale linearly with $n$.\cite{Suleimanov_JPCA:2016} However, since RPMD calculations are approximately $n$ times slower than the purely classical calculations,\cite{Suleimanov_JPCA:2016} after several trials, we have meticulously chosen the number of beads to be 128. This value of $n$ gives a fair compromise between computational cost and accuracy (quantum effects of ZPE and tunneling) at low temperatures. The values of $n$=1, will correspond to a classical calculation.

Since RPMD is simply classical molecular dynamics in an extended ($n$- bead imaginary time path integral) phase space, the ring polymer rate coefficient can be expressed as\cite{Craig_JCP_2:2005,Craig_JCP_1:2005,Suleimanov_JCP:2011}: 

\begin{equation}\label{eq:rpmd_old}
\kRPMD ^{(n)}(T) = \frac{1}{Q_r^{(n)}(T)}\tilde{c}^{(n)}_{fs}(t\rightarrow \infty),
\end{equation}
where, $Q_r^{(n)}(T)$ is the $n$-bead path integral approximation to the quantum mechanical partition function of the reactants per unit volume, and $\tilde{c}^{(n)}_{fs}(t)$ is a ring polymer flux-side correlation function\cite{Miller_JCP:1983}

\begin{equation}
\tilde{c}^{(n)}_{fs}(t) = \frac{1}{(2\pi \hbar)^f}\int d^f \bm{p}_0 \int d^f \bm{r}_0 e^{-\beta H_n(\bm{p}_0,\bm{r}_0)}\delta(\bm{r}_0) v(\bm{r}_0,\bm{p}_0)h(\bm{r}_t).
\end{equation}
Here, subscript 0 and $t$ indicates time, $\delta(\bm{r}_0)$ is a delta function centered at $\bm{r}_0$, $v(\bm{r}_0,\bm{p}_0)$ is the velocity, and $h(\bm{r}_t)$ is the Heaviside step function. The RPMD rate coefficient in \autoref{eq:rpmd_old} is not straightforward to solve numerically. Therefore, we introduce the Bennett-Chandler factorization scheme\cite{Bennett_ACS:1977,Chandler_JCP:1978} to simplify \autoref{eq:rpmd_old} that can be solved numerically without compromising its generality.\cite{Chandler_JCP:1978} The method has been extensively discussed previously\cite{Guevara_JCP:2008,Craig_JCP_1:2005,Guevara_JCP:2009,Suleimanov_JCP:2011} and will not be repeated here. Briefly, in this approach, a reaction coordinate $s(\bm{r})$ is defined, which monitors the progress of a reaction from the reactant ($s > 0$) to the product ($s < 0$) site. For the He transmission through 2D membranes, a reasonable reaction coordinate is $s(\bm{r}) = \bar z$, where $\bar z$ is the $z$- component of the He atom centroid, which follows the minimum energy path of \autoref{fig:PEC}. Within the Bennett-Chandler approach, the RPMD rate coefficient for a process in which reactants and products separated by a dividing surface at $s^{\ddagger}$ (for instance, center of the pore) can be expressed as a product of two terms:

\begin{equation}\label{eq:rpmd_new}
\kRPMD (T) = \kQTST (T) \kappa (t_p).
\end{equation}

Here, the first factor, $\kQTST (T)$, is the centroid-density quantum transition-state theory (QTST) rate coefficient.\cite{Craig_JCP_1:2005} $\kQTST (T)$ can be calculated from the centroid potential of mean force (PMF),\cite{Guevara_JCP:2009,Suleimanov_JCP:2011,Suleimanov_CPC:2013} $W(s)$ along the reaction coordinate. If $s_{\infty}$ is the asymptotic distance in which the He $-$ membrane interaction potential is at the minimum and then introduce a dividing surface carefully placed at the TS region, $s^{\ddagger}$, then $\kQTST (T)$ can be calculated as:\cite{Guevara_JCP:2008,Suleimanov_JPCC:2012}

\begin{equation}\label{eq:kqtst}
\kQTST (T) = \frac{1}{(2\pi\beta m_{\text{He}})^{1/2}}\frac{e^{-\beta W(s^{\ddagger})}}{\int_{s_{\infty}}^{s^{\ddagger}} e^{-\beta W(s)}\dd{s}}.
\end{equation}

$W(s)$'s can be computed by employing the umbrella integration procedure of K\"{a}stner and Thiel.\cite{Kastner_JCP:2005,Kastner_JCP:2006,Kastner_JCP:2009} To calculate the PMF profiles, the reaction coordinate $s$ of the He atom has been divided into 130 equally spaced windows (of width 0.05 \AA) within the range from 6.00 \AA\, (reactant site, $s_{\infty}$) to $-$1.00 \AA\, (product site). 
The PMFs in \autoref{eq:kqtst} is then calculated as:
\begin{equation}\label{eq:pmf}
W(s^{\ddagger}) - W(s_{\infty})= \int_{s_{\infty}}^{s^{\ddagger}} \sum_{i=1}^{N_{\text{windows}}}\left[\frac{N_i P_i(s)}{\sum_{j=1}^{N_{\text{windows}}}N_j P_j(s)}\left(\frac{1}{\beta} \frac{s-\bar{s}_i}{(\sigma_i)^2}-k_i(s-s_i) \right)\right]\dd{s},
\end{equation}

\noindent with 

\begin{equation}
P_i(s) = \frac{1}{\sigma_i\sqrt{2\pi}}\exp\left[-\frac{1}{2}\left(\frac{s-\bar{s}_i}{\sigma_i}\right)^2\right].
\end{equation}

Here, $N_{\text{windows}}$ is the number of biasing windows placed along the reaction coordinate. The strength of the force constant ($k_i$) of the harmonic biasing potential was chosen to be 2.72$\times 10^{-3}T$(K) eV $a_0^{-2}$.  $N_i$ is the total number of steps sampled for window $i$, $\bar{s}_i$ and $\sigma_i^2$ are the mean value and variance for the trajectory calculated for the $i$th window. In each umbrella sampling windows, 100 trajectories with different initial configurations were propagated for 100 ps following an initial equilibration period of 20 ps in the presence of an Andersen thermostat.\cite{Andersen_JCP:1980} The ring polymer equation of motion were integrated using velocity Verlet integrator with a step size of 0.1 fs that involves alternating momentum updates and free ring polymer evolutions.\cite{Craig_JCP:2004}

The second term in \autoref{eq:rpmd_new}, $\kappa (t_p)$, is the long-time limit of a time-dependent ring polymer transmission coefficient or the ring polymer recrossing factor\cite{Markland_JCP:2008,Guevara_JCP:2008,Suleimanov_JCP:2011,Suleimanov_CPC:2013} and is a dynamic correction to $\kQTST$. Typically this factor is calculated at the top of the free energy barrier on the PMF profile so as to minimize the time required to reach plateau value.\cite{Chandler_JCP:1978} This factor ensures that the final $\kRPMD$ value is independent of the choice of the dividing surface.\cite{Craig_JCP_1:2005} The mathematical expression for $\kappa (t_p)$ can be written as:\cite{Suleimanov_JPCC:2012,Suleimanov_CPC:2013}

\begin{equation}\label{eq:kappa}
\kappa (t) = \frac{\expval{\delta [s(\bm{r}_0)]\dot{s}(\bm{r}_0) h[s(\bm{r}_t)]}}{\expval{\delta [s(\bm{r}_0)]\dot{s}(\bm{r}_0) h[\dot{s}(\bm{r}_0)]}},
\end{equation}

\noindent where, $\delta [s(\bm{r}_0)]$ constrains the initial configurations to the dividing surface,  $\dot{s}(\bm{r}_0)$ is the velocity factor that accumulates the flux through the dividing surface, and $h[s(\bm{r}_t)]$is a Heaviside function that gathers trajectories that have crossed over to the product side of the dividing surface. $h[\dot{s}(\bm{r}_0)]$ is basically a normalization factor that ensures $\kappa (t\rightarrow 0_+) = 1$. To calculate the recrossing factor, a long ``parent'' trajectory for He atom of length 2 ns has been carried out after an initial thermalization period of 20 ps in the presence of Andersen thermostat with its centroid pinned at the dividing surface using RATTLE algorithms.\cite{Andersen_JCP:1983} After each 2 ps propagation period of the parent trajectory, 100 trajectories have been generated that have initial position of the parent trajectory, but their momenta is randomly generated from a Boltzmann distribution. These ``child'' trajectories are then propagated for 1 ps in the absence of thermostat and dividing surface constraints. 

\section{Results and discussion}\label{sec:result}

\subsection{Centroid potentials of mean force and ring polymer recrossing factor}

\begin{figure}[!h]
\centering
\includegraphics[width=0.80\textwidth]{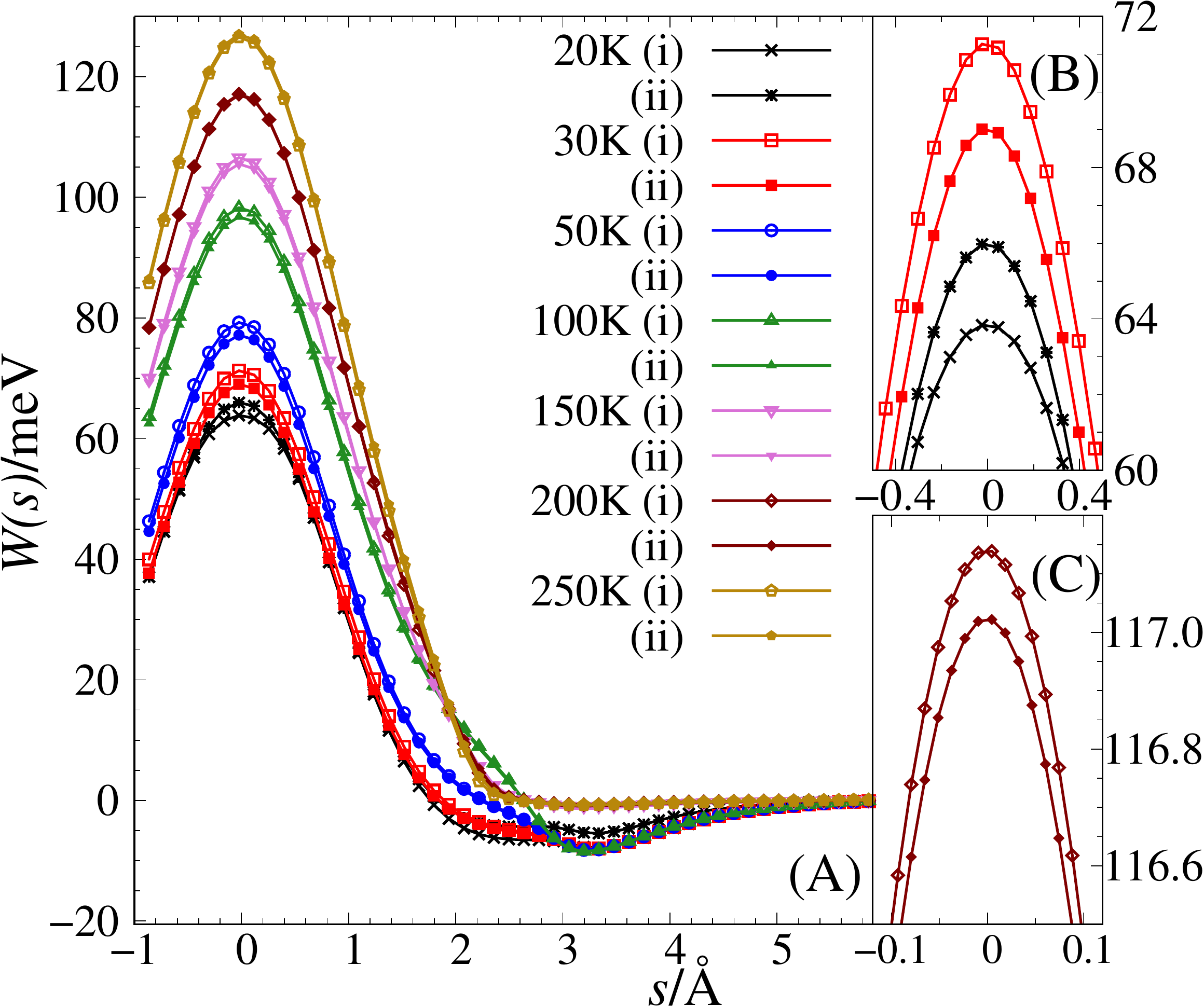} 
\caption{Variation of the RPMD potential of mean force, $W(s)$, (in meV) for (i) $^3$He (ii) $^4$He atom along the reaction coordinate $s$ (in \AA) perpendicular to the graphdiyne membrane within the temperature range (A) 20$-$250 K, (B) 20$-$30 K, and (C) 200 K. The legends correspond to (A), (B), and (C).}
\label{fig:pmf_di}
\end{figure}

The variation of the RPMD potential of mean force $W(s)$ for $^3$He and $^4$He along the reaction coordinate $s$ at $T$ = 20 $-$ 250 K are plotted in \autoref{fig:pmf_di} (Gr2) and \autoref{fig:pmf_tri} (Gr3), and the corresponding $\kQTST$ values are reported in the supplementary information. These PMF profiles include both potential energy and temperature-dependent entropic contributions. The barrier height at the TS, identified around the membrane plane, increases with increasing temperature. The free energy profiles enter a shallow well at around $s$ = 3.0 $-$ 3.4 \AA\, from the reactant site before it steadily increases up to the TS. For the PMFs obtained for the transmission through Gr2, the calculated TS barrier height falls within 64 $-$ 127 meV.  The PMF profiles on Gr3, on the other hand, are more interesting from thermodynamic point of view.  We recall that the He$-$Gr3 interaction potential does not have any barrier;\cite{Marta:2021} rather, a well is built around the transmission zone. This behavior is reflected in the free energy profiles, particularly within the temperature range 20 $-$ 30 K, in which the thermodynamic barrier in the TS region has a negative (or marginally positive) value compared to the reactant site. This energy barrier gradually takes the shape of a parabola with increasing temperature and reaches up to a height of 44 meV at 250 K. 

Since the RPMD method averages points in configurational space on both sides of the potential energy barrier, modeling the effects of the tunneling and ZPE and leaving a mark on the free energy profiles,\cite{Tudela_JPCL:2014} we can roughly attempt to identify these properties. Comparing the PMF profiles of $^4$He at a particular temperature with the corresponding $^3$He's on Gr2 indicate the existence of a prominent ZPE effect; heavier isotope has a lower free energy at the TS than the lighter ones except at 20 K (see \autoref{fig:pmf_di}(B)), in which quantum tunneling may play a crucial role. However, the difference between them at TS decreases with increasing temperature. For example, the difference at 30 K is 2.3 meV (see \autoref{fig:pmf_di}(B)), which decreases to 0.1 meV at 200 K (see \autoref{fig:pmf_di}(C)). Similarly, the $^3$He and $^4$He free energy comparison for Gr3 follows the same trend observed for Gr2, \emph{i.e.}, they decreases with increasing temperature. Within 20 $-$ 30 K, $^4$He has a smaller TS free energy than $^3$He ($\approx$1.1 meV, \autoref{fig:pmf_tri}(B)), and with the increase in temperature in the range 50 $-$ 150 K, the heavier isotope progressively requires more free energy to reach TS, \emph{i.e.}, PMF profiles for both isotope are almost identical. Within 200 $-$ 250 K, the free energies of $^4$He are consistently larger than those obtained for $^3$He (see \autoref{fig:pmf_tri}(C)). These observations can be notionally interpreted as follows: within 20 $-$ 30 K, the free energy barrier is too broadened for effective tunneling, and consequently, the ZPE effect takes precedence.\cite{Tudela_JPCL:2012} There is a competition between them at moderate temperatures (50 $-$ 150 K). At still more elevated temperature (200 $-$ 250 K), the greater tunneling probability of $^3$He may facilitate a smaller TS free energy compared to $^4$He. Note that this analysis is conceptual because these two quantum effects cannot be rigorously separated within the RPMD formalism.

\begin{figure}[!h]
\centering
\includegraphics[width=0.80\textwidth]{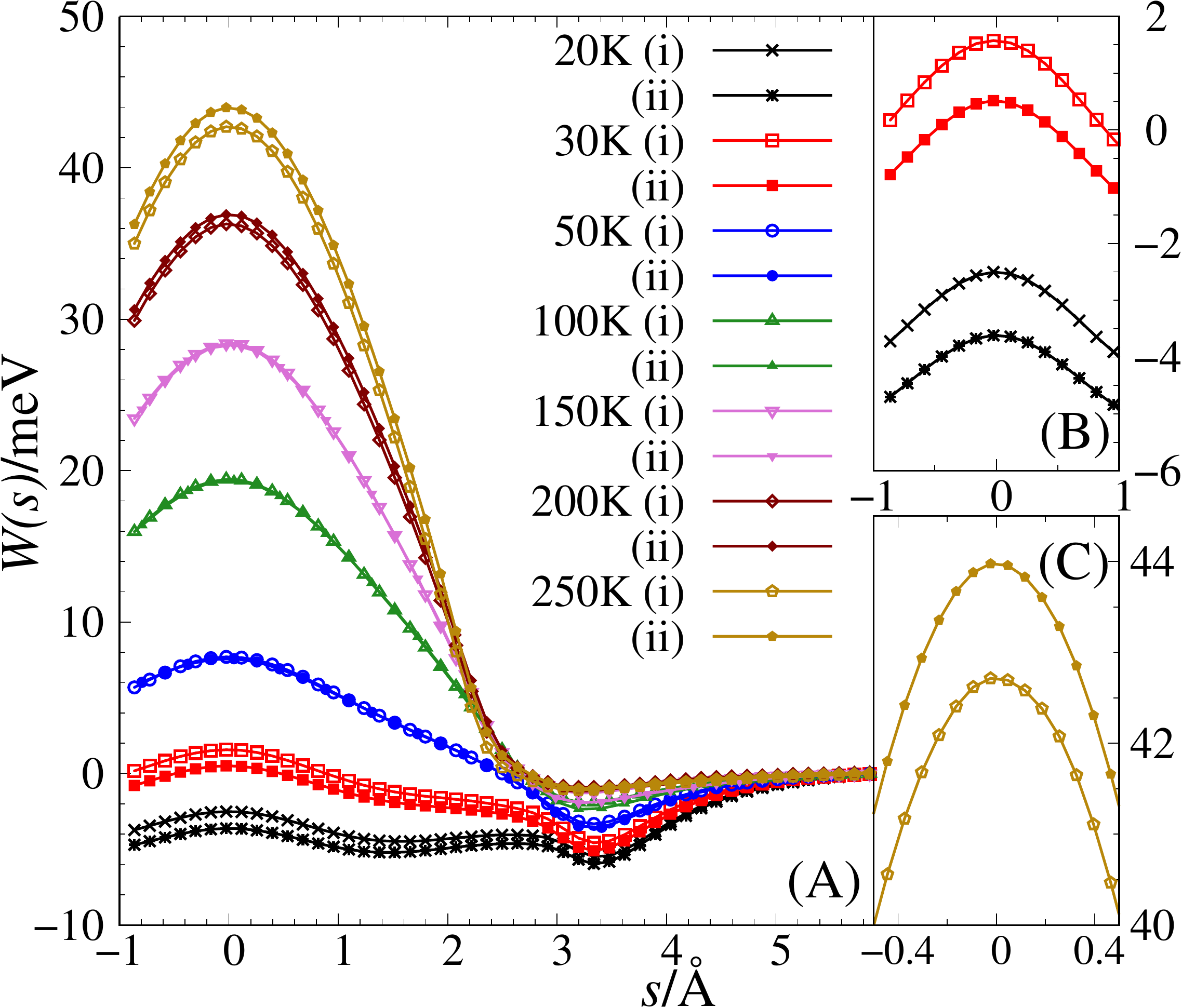} 
\caption{Variation of the RPMD potential of mean force, $W(s)$, (in meV) for (a) $^3$He (b) $^4$He atom along the reaction coordinate $s$ (in \AA) perpendicular to the graphtriyne membrane within the temperature range (A) 20$-$250 K, (B) 20$-$30 K, and (C) 250 K. The legends correspond to (A), (B), and (C).}
\label{fig:pmf_tri}
\end{figure}

There are considerable differences observed between the PMF profiles obtained by the RPMD method with the classical ones at low temperatures (see supplementary information). This, therefore, reinforces our earlier argument of the existence of quantum effects in these transmission processes, which the classical calculation fails to capture. For the He transmission through Gr2 membrane, the free energy barrier height in the classical calculations is always lower than those obtained by the RPMD method. The maximum difference for $^3$He obtained at 100 K ($\approx$ 10 meV) and that for $^4$He obtained at 20 K ($\approx$ 12 meV). Note that the free energy values exponentially contribute to the rate coefficients. As expected, these PMF plots tend to merge with increasing temperature as the quantum effects fade. Comparison between the classical and RPMD method on Gr3 shows that, for both isotopes, the maximum free energy difference at the TS is found at the lowest temperature ($\approx$ 3 meV), and this difference smoothly decreases with increasing temperature. For $^4$He within 200 $-$ 250 K, the difference  between the RPMD and classical barrier heights becomes 1 meV. 

The time-dependent transmission coefficients, $\kappa (t)$, for all temperatures considered in this study on both membranes are always close to unity (0.98 $-$ 1.00) and are moved to the supplementary information (Figure S5 and Figure S6). The corresponding plateau value of the transmission coefficients, $\kappa (t \rightarrow \infty)$, are provided in Table S1. The $\kappa (t)$ value is always close to unity (0.98 $-$ 1.00). Clearly, recrossing dynamics do not play any significant role for the He transmission on both membranes. Particularly, from 50 K and higher temperature, where $\kappa (t)$ value is always 1.00. This implies that most of the He trajectories that reach the center of the pore of either Gr2 or Gr3 would overcome the free energy barrier and transport to the other side of the dividing surface. At 20 K, $\kappa (t)$ for $^3$He is marginally smaller than that obtained for $^4$He (0.98 and 0.99 respectively). Note that the classical recrossing factor is always 1, even at low temperatures.

\subsection{Thermal rate coefficient}

Since the plateau of $\kappa (t)$ is always close to unity, there is practically no significant difference between the values of $\kQTST$ and $\kRPMD$ on both membranes. The variation of $\kRPMD$ with temperature is plotted in \autoref{fig:rate} and Figure S7 of the supplementary material, while the numerical values of $\kRPMD$ and $\kQTST$ are reported in Table S1 (see supplementary information). It is evident that $\kRPMD$'s on Gr3 is many orders of magnitude greater than that obtained on Gr2. The maximum difference was observed at the lowest temperature (by a factor of 10$^{17}$ at 20 K), and with the increase in temperature, this difference decreases rapidly (by a factor of 10$^{2}$ at 250 K) as the $\kRPMD$ vs. $T$ curves start to converge. This is due to the fact that the value of $\kRPMD$ on Gr3 does not change drastically with temperature (confined within 3.63$\times$10$^{10}$ s$^{-1}$ $-$ 1.01$\times$10$^{11}$ s$^{-1}$). It starts to decrease slightly with temperature up to 50 K and then increases successively with increasing $T$. On Gr2, however, for both isotopes, $\kRPMD$ increases manifold with temperature, in agreement with the previous calculations. \cite{Bartolomei_JPCC:2014,Hernandez_JPCA:2015,Hernandez_JPCC:2017} This increment in the rate coefficient is more pronounced at the modest rise in $T$ in the low temperature regime than at the high temperature range. For example, the increase in the value of $\kRPMD$ is in the order of 10$^{5}$ for the temperature rise from 20 K to 30 K or 30 K to 50 K. However, $\kRPMD$ increases much less than an order for the consecutive 50 K temperature jumps, starting at 150 K. 

\begin{figure}[!h]
   \centering
\includegraphics[width=0.70\textwidth]{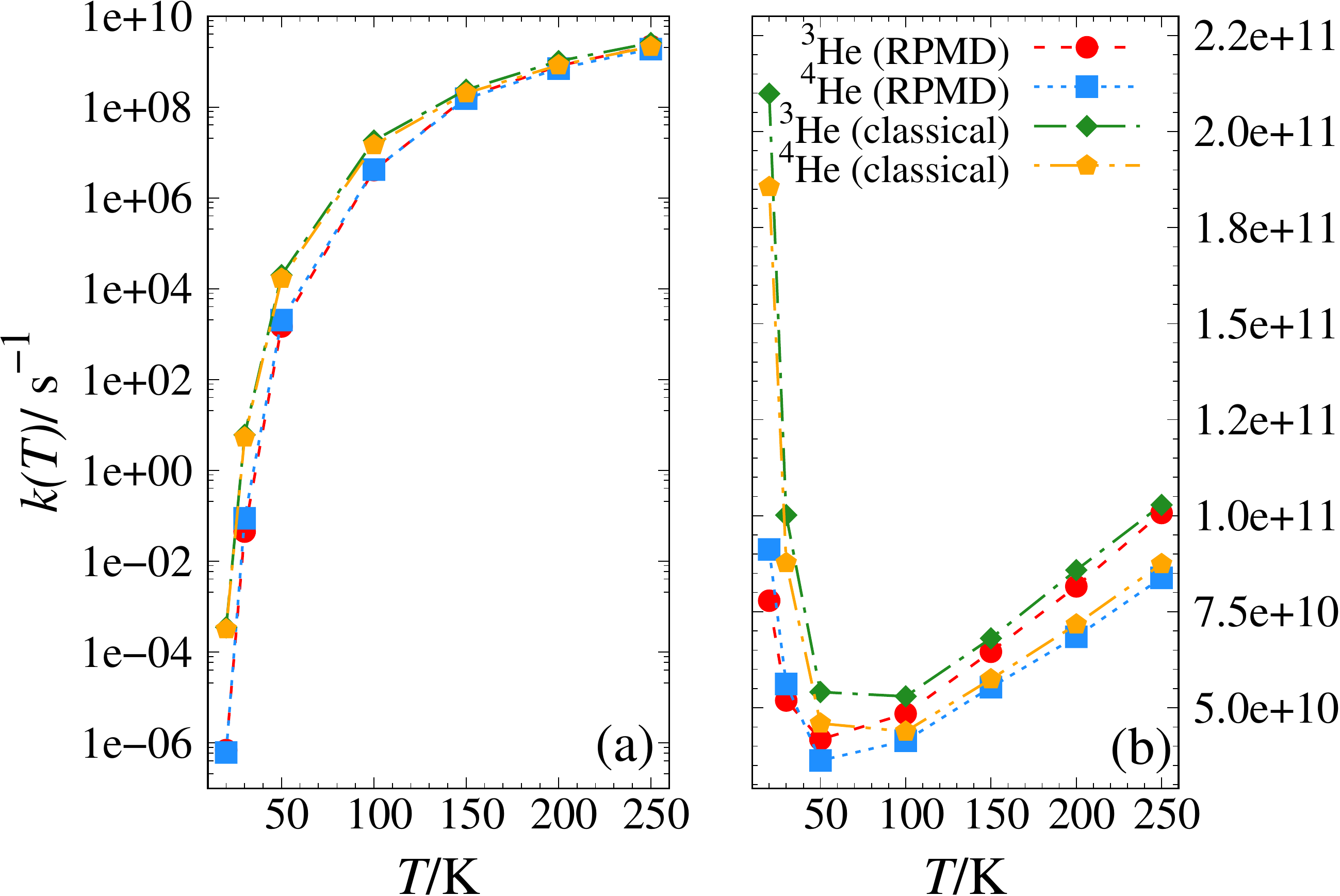} 
 \caption{Variation of the ring polymer molecular dynamics, RPMD ($\kRPMD$) and classical ($\kcl$) rate coefficients  (in s$^{-1}$) for the transmission of $^3$He [red circle (RPMD) and green diamond (classical)] and  $^4$He [blue square (RPMD) and orange pentagon (classical)] through (a) graphdiyne and (b) graphtriyne membranes with temperature $T$ (in K). The legends correspond to both (a) and (b).}
\label{fig:rate}
\end{figure}

The above observations can be explained by inspecting the corresponding PMF profiles. At low temperatures (20 K $-$ 50 K), there is practically no free energy barrier for the transmission of either isotope on Gr3. However, on Gr2, the free energy barrier is already above 60 meV at 20 K. The slowly moving He atoms do not have enough kinetic energy to overcome this barrier to reach the other side of the membrane. Therefore, it is not surprising that $\kRPMD$'s on Gr3 is many orders of magnitude greater than those obtained on Gr2, and the actual He flux through the pores of Gr2 membranes will be extremely slow.\cite{Hernandez_JPCA:2015,Hernandez_JPCC:2017} Similar arguments can be presented at higher temperatures, although the increased kinetic energy of the incoming He atom will contribute to smaller differences in the $\kRPMD$ values. We also point out the analogous quantum wave packet observation reported on the Gr2 and holey graphene (P7) sheet having a more diffused pore.\cite{Hernandez_JPCC:2017}  
Finally, when comparing the classical and RPMD rate coefficients, it is obvious that the classical method overestimates the rate coefficient within the whole temperature regime studied in this work. This discrepancy is particularly more apparent at low temperatures, when the quantum effects dominate, and on the Gr2 membrane. For example, at 20 K and for Gr2 membrane, the classical rate coefficient is more than three orders of magnitude greater than the corresponding $\kRPMD$.  However, they do closely follow a similar temperature dependence as obtained by the RPMD calculations.

\subsection{$^4$He/$^3$He selectivity}

The $^4$He/$^3$He selectivity, defined as the ratio between the RPMD rate coefficient of the heavier He isotope to the lighter one, $\kRPMD$($^4$He)/$\kRPMD$($^3$He), is plotted as a function of temperature in \autoref{fig:ratio}, and the corresponding values are reported in Table S1 (see supplementary information). For Gr2 membrane, $^4$He/$^3$He selectivity increases for the temperature rise from 20 K to 30 K. At 30 K, the maximum selectivity is obtained ($\approx$ 2) favouring the heavier isotope and simultaneously indicating pronounced quantum effects as discussed previously. However, this increased selectivity should be accompanied by a decreased permeability.\cite{Robeson_JMS:2008} With the increase in temperature, this selectivity ratio progressively becomes smaller and starts to flatten out, commencing from 100 K. Although at higher temperatures ($>$ 100 K), the transmission of the lighter isotope is favoured to some extent. We note the analogous selectivity profile was obtained on the P7 sheet.\cite{Hernandez_JPCC:2017} Similarly, for the Gr3 membrane, the maximum $^4$He/$^3$He selectivity obtained at 20 K (1.17), which gradually decreases till 50 K, and then remains almost constant (0.87 to 0.83) marginally preferring $^3$He for higher temperature regime. The variation of the selectivity is fully consistent with the nature of their PMF profiles that inherits quantum effects. It is interesting to note that the selectivity plot for both Gr2 and Gr3 approach each other with increasing $T$, one from the top and another from the bottom, and at 250 K, they almost become equal ($\approx$ 0.8 and partially in favour of $^3$He). This is reasonable since as the quantum effects diminish with temperature and with fewer constraints, the lighter isotope transmission will be slightly favoured. 

\begin{figure}[!h]
   \centering
\includegraphics[width=0.70\textwidth]{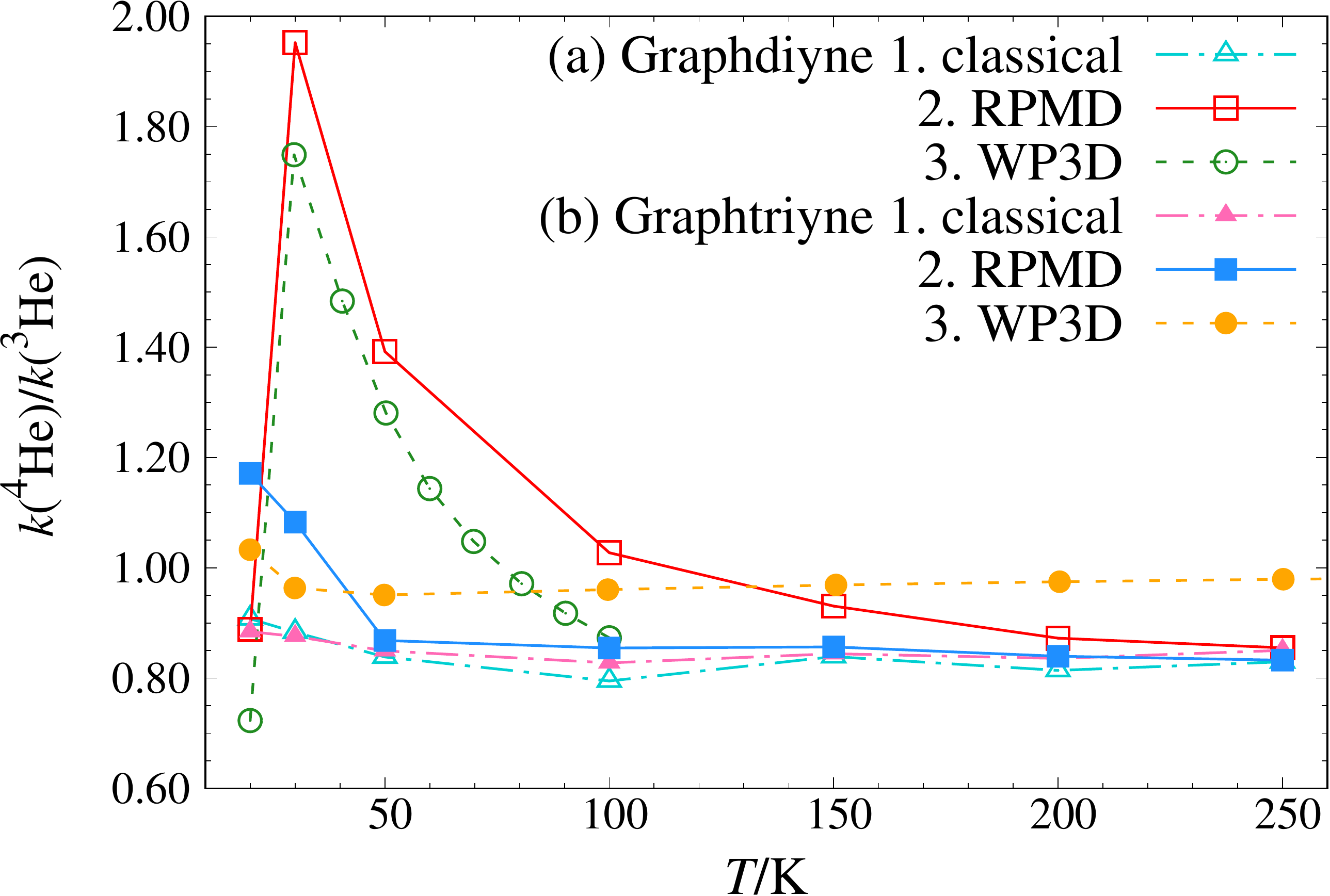} 
    \caption{Variation of the $^4$He/$^3$He rate coefficient ratio, $k$($^4$He)/$k$($^3$He), calculated using 1. classical (triangle) 2. ring polymer molecular dynamics, RPMD (square) and 3. three-dimensional wave packet propagation method, WP3D (circle) for the He transmission through (a) graphdiyne and (b) graphtriyne membranes with temperature $T$ (in K).}
\label{fig:ratio}
\end{figure}

The selectivity ratios obtained by the RPMD method show good agreement with those obtained by the quantum three-dimensional wave packet propagation (WP3D) results obtained by Gij{\'{o}}n \emph{et al.}\cite{Hernandez_JPCC:2017} and Hern{\'a}ndez \emph{et al.}\cite{Marta:2021}  for the whole temperature range studied in this work. In general, the RPMD selectivity profile closely resembles the WP3D results; however, they do overestimate and underestimate to some degree. For example, on Gr2, the RPMD selectivity ratio overestimates the quantum calculations for the whole temperature range. The maximum difference between these two methods is found at 30 K (0.2). Similarly, on Gr3, the RPMD method seems to overestimate the selectivity ratio within 20 K $-$ 30 K and underestimate them for higher temperatures by a maximum of 0.15. Overall, the difference between RPMD and WP3D results for the selectivity ratio lies within 9$-$23\% on Gr2 and 9$-$15\% on Gr3 in comparison to the quantum calculation. On the other hand, the classical selectivity shows no significant change with temperature and remains almost constant (0.91 $-$ 0.81), slightly favoring $^3$He transmission. This is valid for both membranes. As expected, with the rise in temperature, the classical selectivity follows the RPMD and WP3D ones.

\section{Concluding remarks}\label{sec:conclusion}

In this work, we have calculated the thermal rate coefficient for the transmission of He isotopes through the pores of one atom thick graphdiyne (Gr2) and graphtriyne (Gr3) membranes using the ring polymer molecular dynamics (RPMD) method within the temperature range 20 K $-$ 250 K. Transmission through Gr2 has a substantial free energy barrier even at 20 K, and this barrier height increases with increasing temperature. On the other hand, transmission through Gr3 can either have marginally negative (up to 30 K) or small positive (T$\ge$ 50 K) free energy barrier. The extent of the barrier height directly impacts the calculated rates, as evident from the fact that the rate coefficient on Gr3 is at least 10$^{17}$ order of magnitude greater than on Gr2 at 20 K. The rate coefficient on Gr2 increases rapidly with temperature and starts to converge, starting from 150 K ($\sim$ 10$^8$ s$^{-1}$). However, the rate coefficients on Gr3 do not vary appreciably with temperature and remain almost constant for the whole temperature range (3.63$\times$10$^{10}$ s$^{-1}$ $-$ 1.01$\times$10$^{11}$ s$^{-1}$). In general, the rate coefficient calculated for the transmission through Gr3 is always greater than the corresponding one on Gr2 over the whole temperature regime considered in this work. Moreover, we found that the recrossing dynamics have little or no effect on the final value of the rate coefficients.

The selectivity ratio, which indicates the preference of either isotope for its permeation through the membranes, has been calculated as a function of temperature. From the values of the selectivity ratio, the quantum effects driving the He transmission, particularly at low temperatures, \emph{i.e.}, the zero-point energy (ZPE) favoring the heavier isotope and the tunneling of the more mobile lighter isotope, has been coarsely deduced. The maximum selectivity ratio found in this study ($\approx$ 2) was obtained on Gr2 membrane at 30 K in which the rate of transmission of $^4$He is almost twice the rate of $^3$He seemingly due to a more dominant ZPE effect. A similar conclusion can be derived on Gr3 at 20 K, where the $^4$He/$^3$He selectivity is around 1.2. However, on the Gr3 membrane, the selectivity ratio does not vary considerably with the temperature (0.8$-$1.2) as compared to what was observed for the transmission through the Gr2 membrane (0.9$-$2.0). With the increase in temperature, the permeation of the lighter isotope is marginally favored. The RPMD selectivity ratio is consistent with the quantum calculations for the entire temperature range studied in this work.\cite{Hernandez_JPCC:2017} On the other hand, the classical method failing to capture either of the quantum effects demonstrated a considerable discrepancy with the RPMD and quantum results, particularly at low temperatures. Therefore, it is of paramount importance to use robust and accurate methods, such as RPMD, that can correctly describe quantum effects such as ZPE and tunneling  when studying physical processes for which strong quantum nature is expected.

In conclusion, in the present study, we have corroborated the efficient and rigorous nature of the RPMD method by determining thermal rate coefficients of physical processes of broad industrial and scientific significance. We also hope that this work will stimulate future experimental measurements of the rate coefficients for the separation of He isotopes using graphene derivatives, taking advantage of quantum effects at low temperatures. As an extension to this work, we would like to investigate the influence of surrounding He atoms on the rate coefficient and apply the RPMD method for isotopic He separation on other graphene derivatives such as polyphenylene, functionalized graphene pores, holey graphene, graphenylene membranes, \emph{etc}. that were previously reported to serve as excellent atomic sieves.

\begin{acknowledgement}

Y.V.S. and S.B. acknowledge the support of the European Regional Development Fund and the Republic of Cyprus through the Research Promotion Foundation (Projects: INFRASTRUCTURE/1216/0070 and Cy-Tera NEA ${\rm Y\Pi O\Delta OMH}$/${\rm \Sigma TPATH}$/0308/31). Y.V.S. was also supported by RFBR grant number 20-03-00833. J.C.-M. and M.I.H. were supported by Spanish MICINN with Grant FIS2017-84391-C2-2-P.  We are also indebted to ``Centro de Supercomputaci\'on de Galicia'' (CESGA) and specially to the people in the help desk.

\end{acknowledgement}

\begin{suppinfo}

The following files are available free of charge.
\begin{itemize}
  \item supplementary information file: computational details and additional results.
\end{itemize}

\end{suppinfo}

\bibliography{He-Graph}

\end{document}

\section{Potential of mean force}

\begin{figure}[H]
   \centering
   \includegraphics[width=0.70\textwidth]{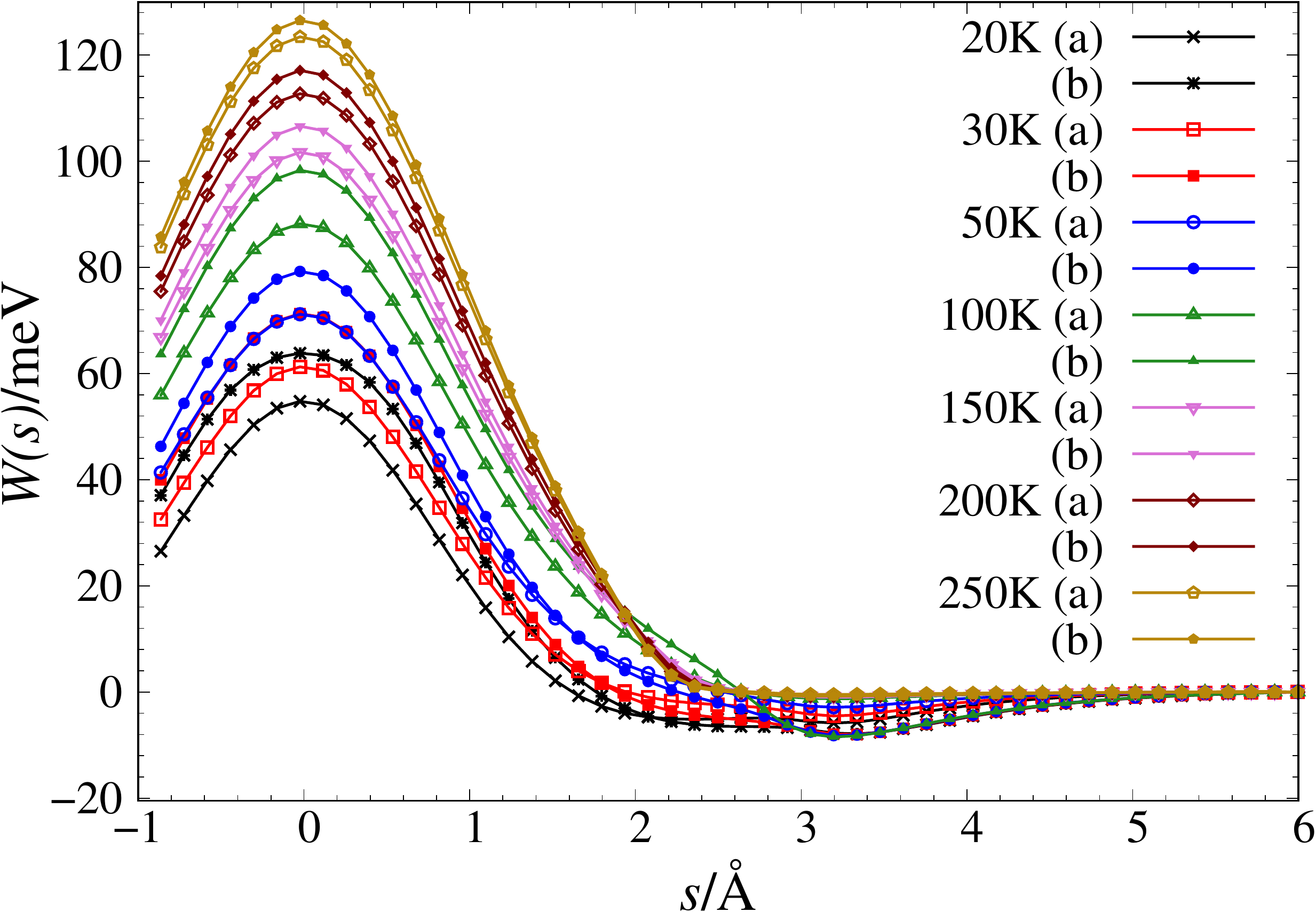} 
   \caption{Variation of the (a) classical and (b) RPMD potential of mean force, $W(s)$, (in meV) for $^3$He atom along the reaction coordinate $s$ (in \AA) perpendicular to the graphdiyne membrane within the temperature range 20$-$250 K.}
\end{figure}

\begin{figure}[H]
   \centering
   \includegraphics[width=0.70\textwidth]{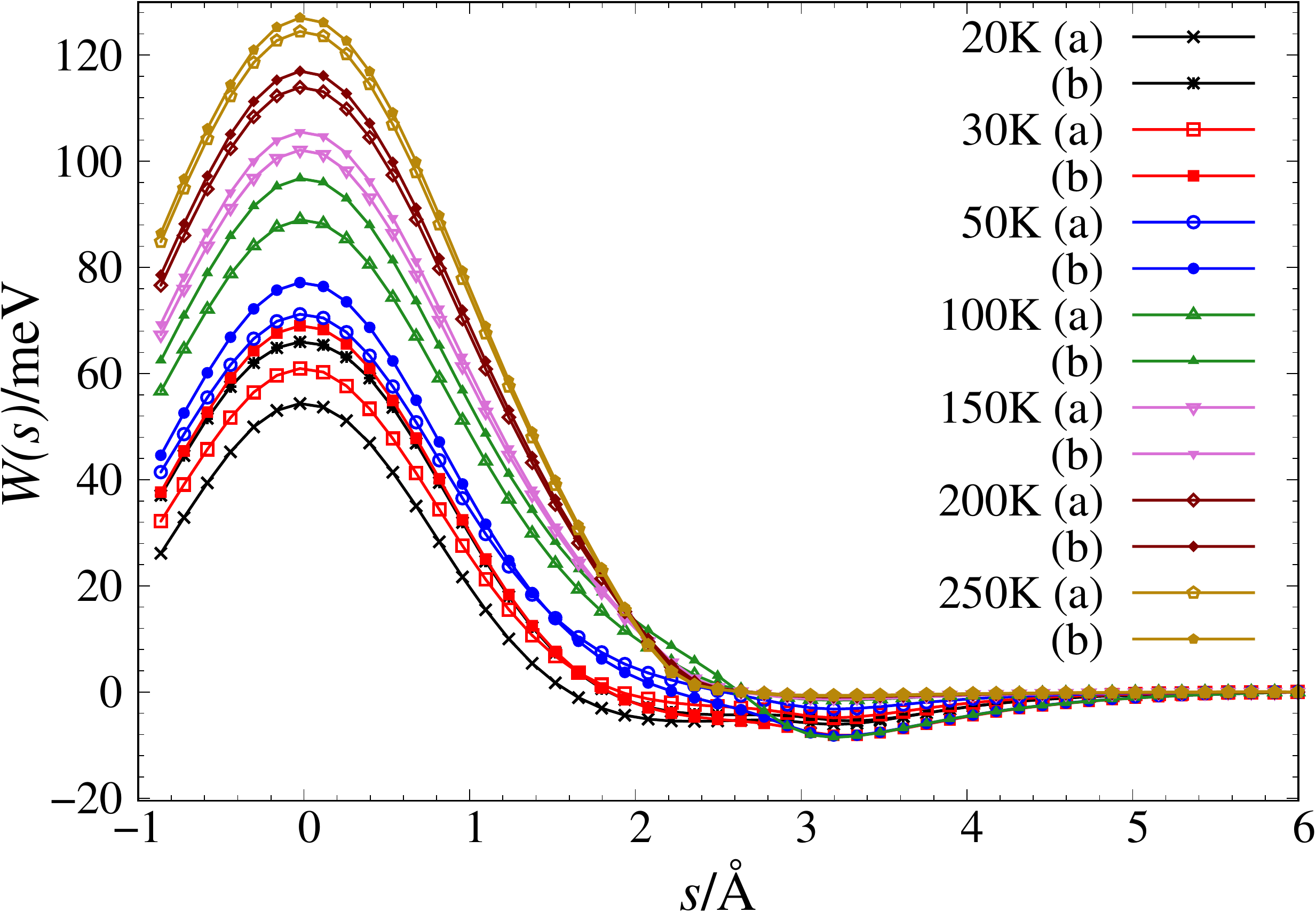} 
   \caption{Variation of the (a) classical and (b) RPMD potential of mean force, $W(s)$, (in meV) for $^4$He atom along the reaction coordinate $s$ (in \AA) perpendicular to the graphdiyne membrane within the temperature range 20$-$250 K.}
\end{figure}

\begin{figure}[H]
   \centering
   \includegraphics[width=0.70\textwidth]{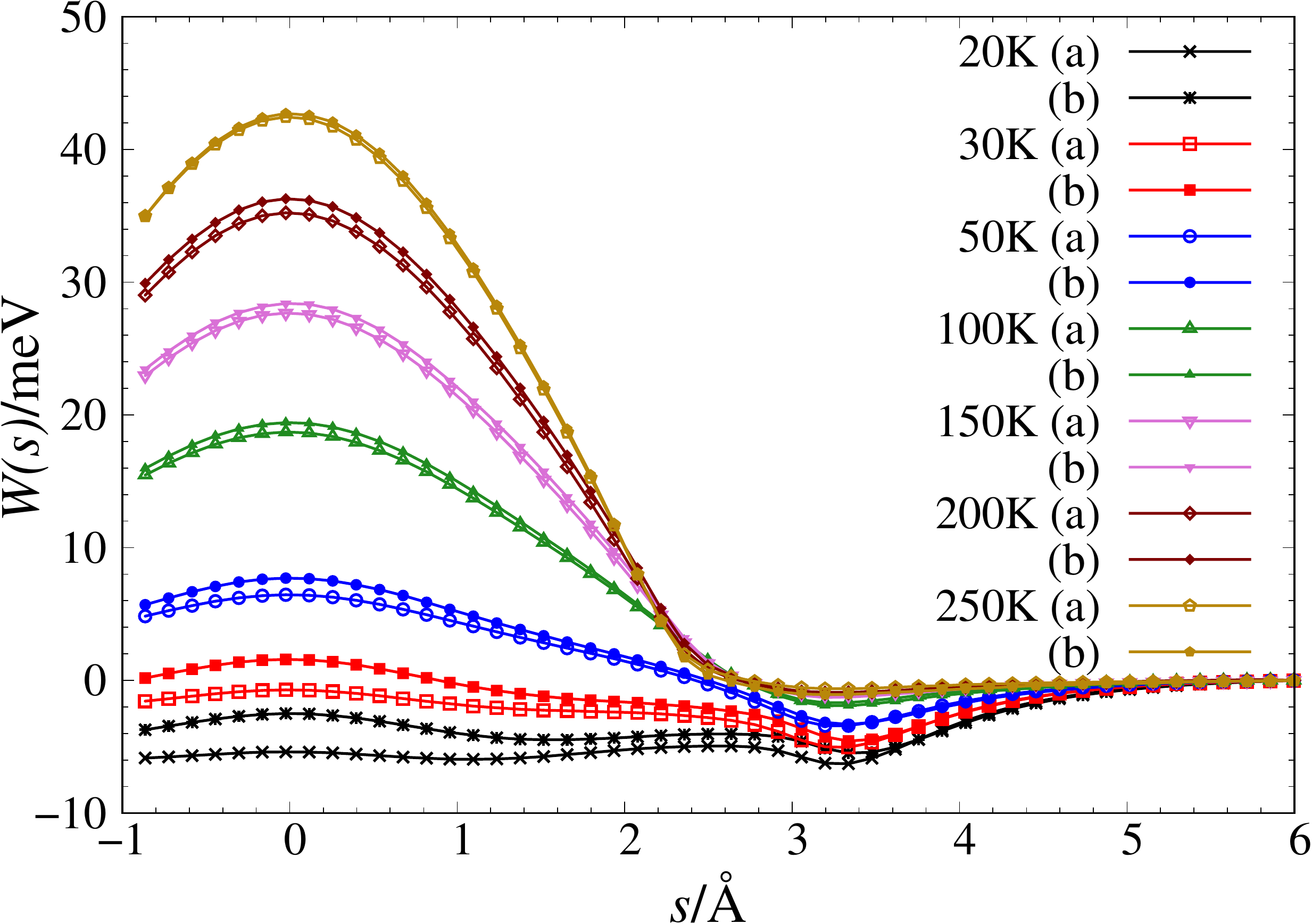} 
   \caption{Variation of the (a) classical and (b) RPMD potential of mean force, $W(s)$, (in meV) for $^3$He atom along the reaction coordinate $s$ (in \AA) perpendicular to the graphtriyne membrane within the temperature range 20$-$250 K.}
\end{figure}

\begin{figure}[H]
   \centering
   \includegraphics[width=0.70\textwidth]{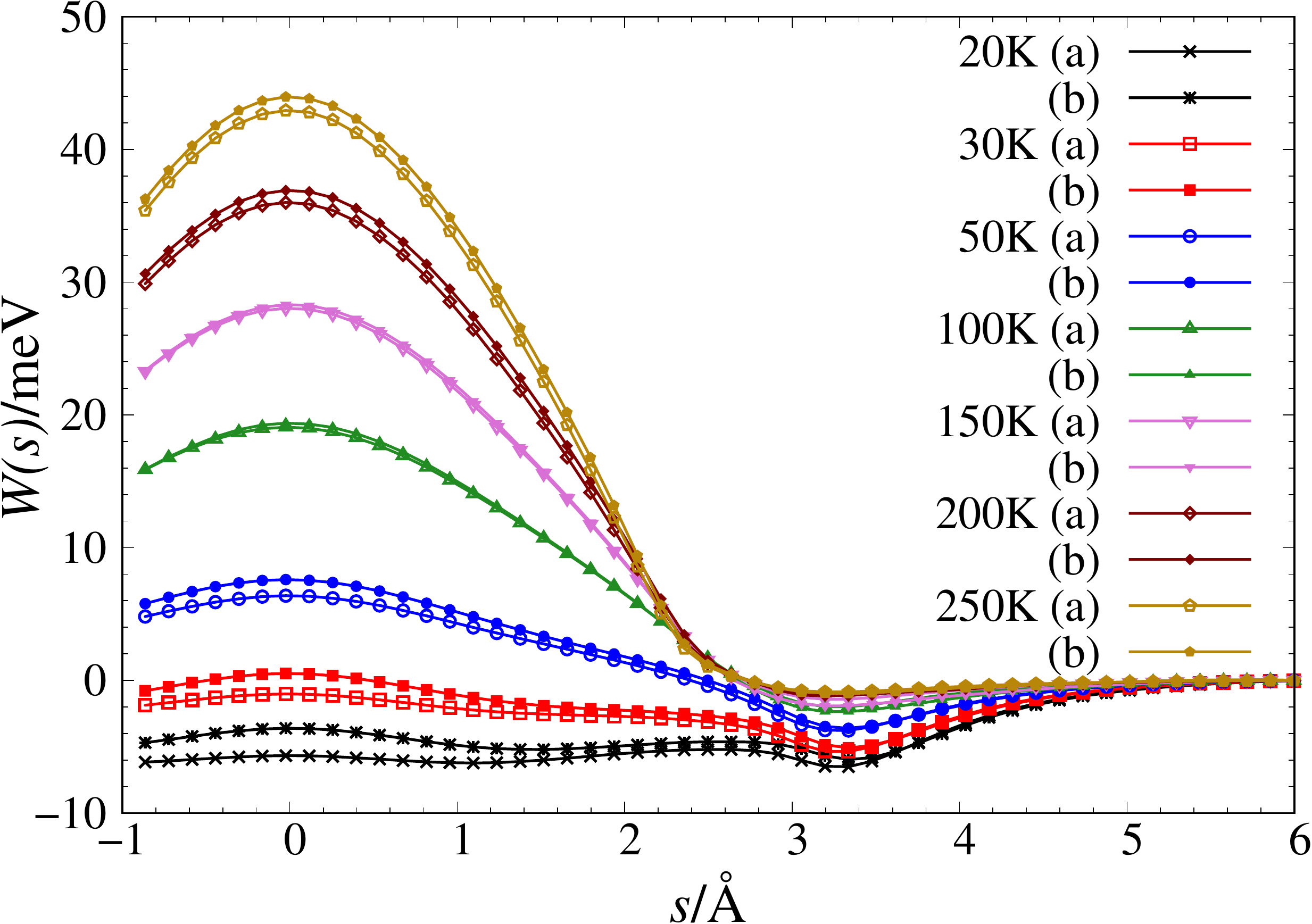} 
   \caption{Variation of the (a) classical and (b) RPMD potential of mean force, $W(s)$, (in meV) for $^4$He atom along the reaction coordinate $s$ (in \AA) perpendicular to the graphtriyne membrane within the temperature range 20$-$250 K.}
\end{figure}

\section{Ring polymer transmission coefficient}

\begin{figure}[H]
   \centering
   \includegraphics[width=0.70\textwidth]{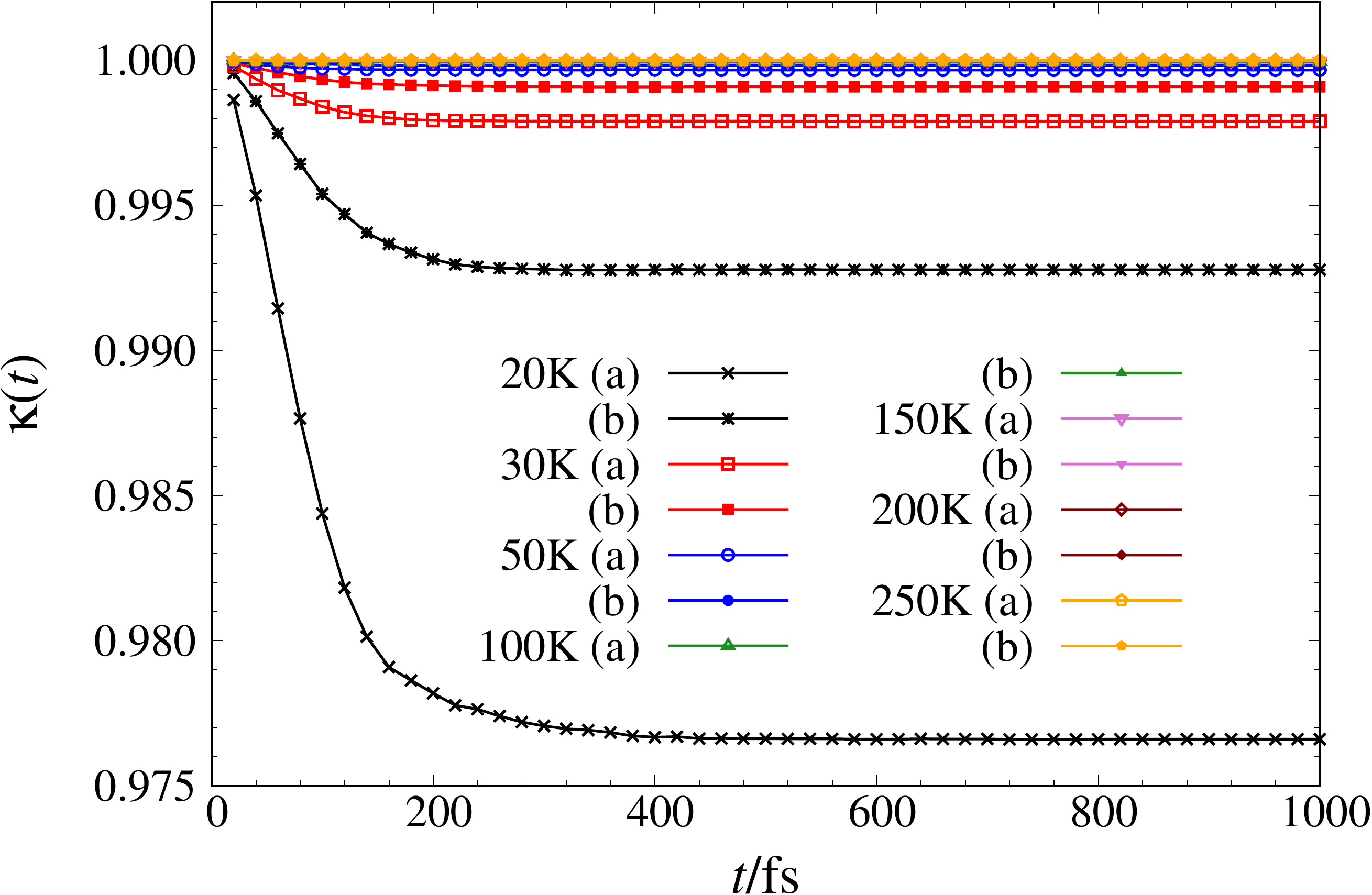} 
   \caption{Ring polymer time dependent transmission coefficient, $\kappa (t)$, in the temperature range 20$-$250 K for (a) $^3$He and (b) $^4$He atom transmission through the graphdiyne membrane.}
\end{figure}

\begin{figure}[H]
   \centering
   \includegraphics[width=0.70\textwidth]{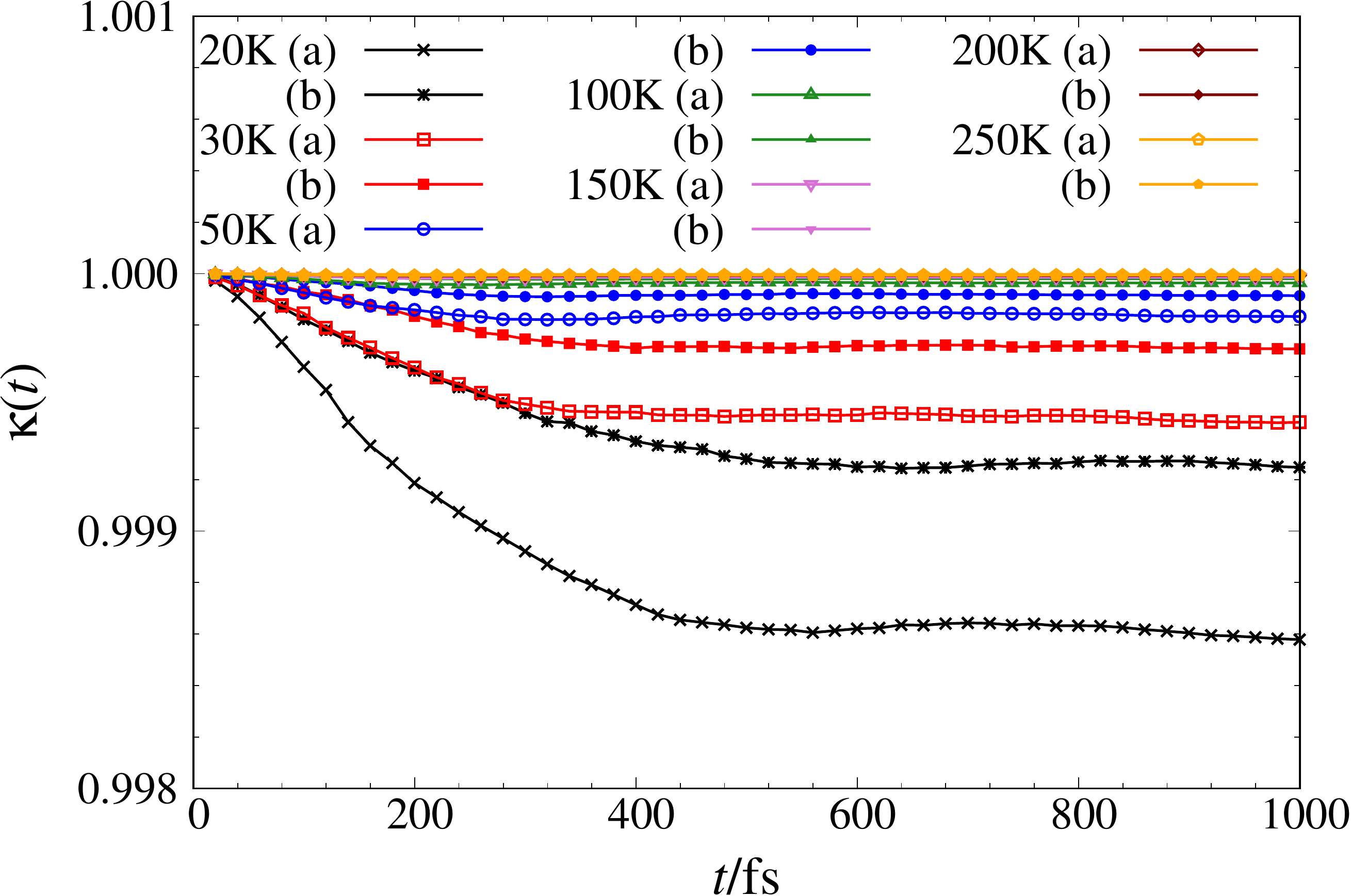} 
   \caption{Ring polymer time dependent transmission coefficient, $\kappa (t)$, in the temperature range 20$-$250 K for (a) $^3$He and (b) $^4$He atom transmission through the graphtriyne membrane.}
\end{figure}

\section{Rate coefficient and selectivity}

\begin{figure}[H]
   \centering
\includegraphics[width=0.70\textwidth]{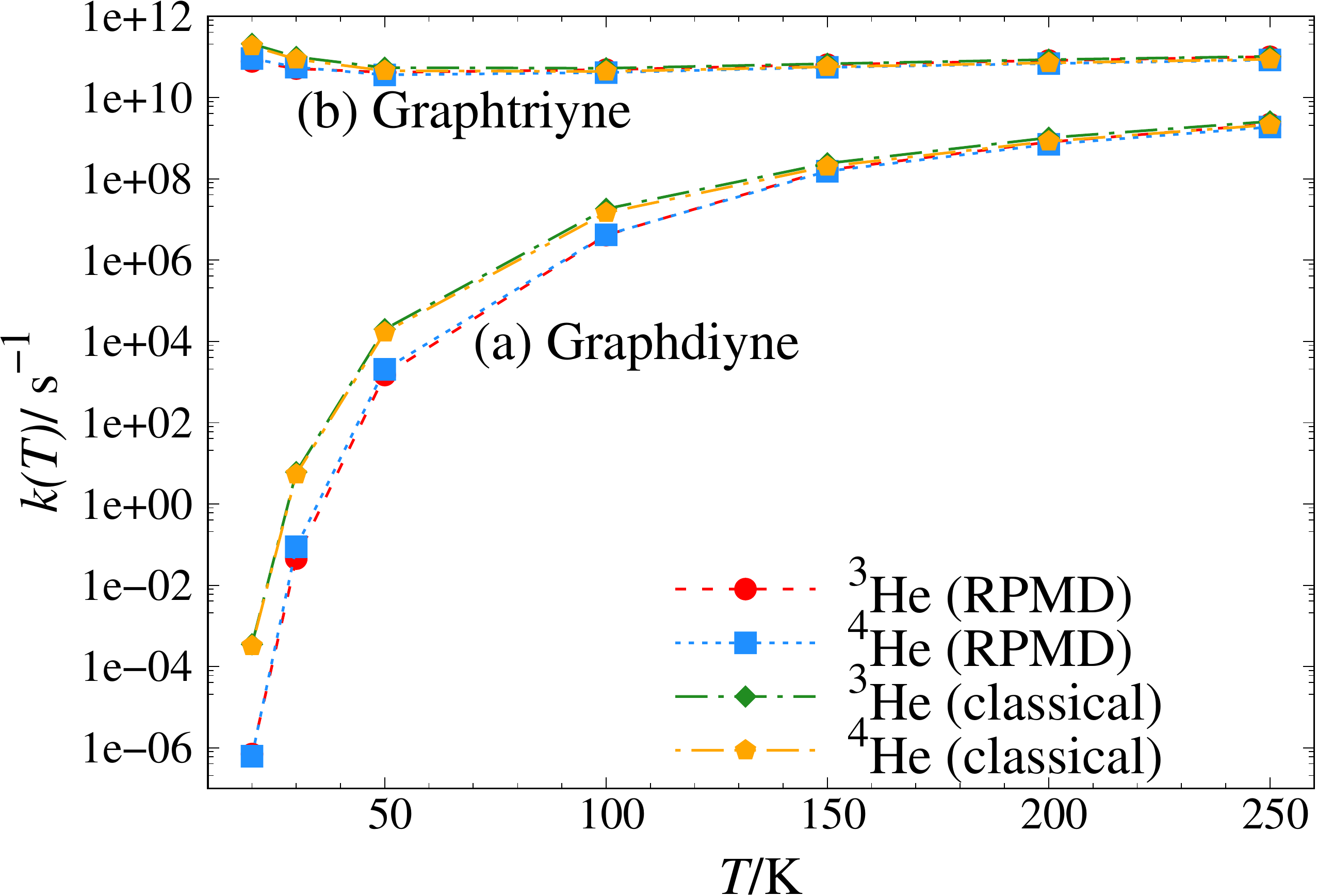} 
   \caption{Variation of the ring polymer molecular dynamics, RPMD ($\kRPMD$) and classical ($\kcl$) rate coefficients  (in s$^{-1}$) for the transmission of $^3$He [red circle (RPMD) and green diamond (classical)] and  $^4$He [blue square (RPMD) and orange pentagon (classical)] through (a) graphdiyne and (b) graphtriyne membranes with temperature $T$ (in K).}
\label{fig:rate}
\end{figure}

\begin{table}[H]
\caption{Summary of the rate calculations for $^3$He and $^4$He atom transmission through the graphdiyne (Gr2) and graphtriyne (Gr3) membranes at temperatures ($T$) 20, 30, 50, 100, 150, 200, and 100 K: $\kQTST$ ($\kTST$) $-$ centroid-density quantum (classical) transition state theory rate coefficient;$\textsuperscript{\emph{a}}$ $\kappaRPMD$ ($\kappacl$) $-$ ring polymer (classical) transmission coefficient; $\kRPMD$ ($\kcl$) $-$ ring polymer (classical) molecular dynamics rate coefficient;$\textsuperscript{\emph{a}}$ $^4$He/$^3$He $-$ ratio between the $^4$He and $^3$He rate coefficient.}
\small
\begin{center}
\begin{tabular}{lll|lllr|lllr}
\hline
\multirow{2}{*}{} & \multirow{2}{*}{$T$/K} & \multirow{2}{*}{Isotope} & \multicolumn{4}{c}{classical} & \multicolumn{4}{c}{RPMD} \\
\cline{4-7}
\cline{8-11}
&&&$\kTST$&$\kappacl$&$\kcl$ & $^4$He/$^3$He&$\kQTST$&$\kappaRPMD$&$\kRPMD$ & $^4$He/$^3$He\\
\hline
\hline
\multirow{14}{*}{Gr2} & \multirow{2}{*}{20} & $^3$He & 3.54(-4)&1.00 &3.54(-4)&\multirow{2}{*}{0.91}&7.15(-7) & 0.98 & 6.99(-7) &\multirow{2}{*}{0.89}\\
&&$^4$He & 3.21(-4) & 1.00 &3.21(-4)&&6.25(-7) & 0.99 & 6.21(-7)\\
\cline{2-11}
& \multirow{2}{*}{30} & $^3$He &6.09(0) &1.00&6.09(0)&\multirow{2}{*}{0.88} & 4.53(-2) & 0.99 & 4.52(-2)&\multirow{2}{*}{1.95}\\
&&$^4$He & 5.38(0) &1.00 & 5.38(0) && 8.84(-2) & 0.99 & 8.83(-2)\\
\cline{2-11}
& \multirow{2}{*}{50} & $^3$He &2.00(4) &1.00&2.00(4)&\multirow{2}{*}{0.84}&1.48(3) &1.00 & 1.48(3) &\multirow{2}{*}{1.39}\\
&&$^4$He & 1.68(4) & 1.00 & 1.68(4) &&2.06(3) & 1.00 & 2.06(3)\\
\cline{2-11}
& \multirow{2}{*}{100} & $^3$He & 1.84(7) &1.00 & 1.84(7) &\multirow{2}{*}{0.80} & 4.10(6) & 1.00 & 4.10(6) &\multirow{2}{*}{1.03}\\
&&$^4$He & 1.46(7) & 1.00 & 1.46(7) &&4.21(6) & 1.00 & 4.21(6)\\
\cline{2-11}
& \multirow{2}{*}{150} & $^3$He & 2.41(8) &1.00 & 2.41(8) &\multirow{2}{*}{0.84} & 1.65(8) & 1.00 & 1.65(8) &\multirow{2}{*}{0.93}\\
&&$^4$He & 2.02(8) &1.00 &2.02(8)&&1.54(8) & 1.00 & 1.54(8)\\
\cline{2-11}
& \multirow{2}{*}{200} & $^3$He & 1.03(9) & 1.00 & 1.03(9) &\multirow{2}{*}{0.81} & 8.07(8) & 1.00 & 8.07(8) &\multirow{2}{*}{0.87}\\
&&$^4$He & 8.40(8) &1.00&8.40(8)&&7.04(8) & 1.00 & 7.04(8)\\
\cline{2-11}
& \multirow{2}{*}{250} & $^3$He & 2.57(9) & 1.00 &2.57(9) &\multirow{2}{*}{0.83} & 2.22(9) & 1.00 & 2.22(9) &\multirow{2}{*}{0.85}\\
&&$^4$He & 2.14(9) &1.00&2.14(9)&&1.90(9) & 1.00 & 1.90(9)\\
\hline
\hline
\multirow{14}{*}{Gr3} & \multirow{2}{*}{20} & $^3$He & 2.10(11) & 1.00 & 2.10(11) &\multirow{2}{*}{0.88}&7.80(10) &0.99&7.79(10)& \multirow{2}{*}{1.17}\\
&&$^4$He & 1.86(11) & 1.00 & 1.86(11) && 9.13(10) & 0.99 & 9.12(10)\\
\cline{2-11}
& \multirow{2}{*}{30} & $^3$He &1.00(11) & 1.00 & 1.00(11) &\multirow{2}{*}{0.88}& 5.19(10) &1.00 & 5.19(10)&\multirow{2}{*}{1.08}\\
&&$^4$He &8.77(10)  &1.00 &8.77(10) && 5.62(10) & 1.00 & 5.62(10)\\
\cline{2-11}
& \multirow{2}{*}{50} & $^3$He &5.41(10) & 1.00 & 5.41(10) &\multirow{2}{*}{0.85}&4.18(10) &1.00 & 4.18(10)&\multirow{2}{*}{0.87}\\
&&$^4$He & 4.60(10) & 1.00 & 4.60(10) &&3.63(10) & 1.00 & 3.63(10)\\
\cline{2-11}
& \multirow{2}{*}{100} & $^3$He & 5.29(10)&1.00&5.29(10)&\multirow{2}{*}{0.83}&4.84(10) &1.00  & 4.84(10) &\multirow{2}{*}{0.86}\\
&&$^4$He & 4.38(10) &1.00 & 4.38(10) &&4.14(10) & 1.00 & 4.14(10)\\
\cline{2-11}
& \multirow{2}{*}{150} & $^3$He & 6.81(10) & 1.00 &6.81(10)&\multirow{2}{*}{0.84} &6.46(10) & 1.00 & 6.46(10)&\multirow{2}{*}{0.86}\\
&&$^4$He & 5.74(10) & 1.00 & 5.74(10) &&5.53(10) & 1.00 & 5.53(10)\\
\cline{2-11}
& \multirow{2}{*}{200} & $^3$He & 8.58(10) & 1.00 & 8.58(10) &\multirow{2}{*}{0.84} &8.16(10) &1.00 &8.16(10)&\multirow{2}{*}{0.84}\\
&&$^4$He & 7.17(10) &1.00 & 7.17(10) &&6.85(10)& 1.00 & 6.85(10)\\
\cline{2-11}
& \multirow{2}{*}{250} & $^3$He & 1.03(11) & 1.00 &1.03(11)&\multirow{2}{*}{0.85} &1.01(11) &1.00 & 1.01(11) &\multirow{2}{*}{0.83}\\
&&$^4$He & 8.74(10) &1.00 & 8.74(10) &&8.38(10) & 1.00 & 8.38(10)\\
\hline
\hline
\end{tabular}%
\end{center}
 \textsuperscript{\emph{a}} The thermal coefficients are given in s$^{-1}$, and the numbers in the parentheses denote powers of ten.\\
\label{tab:result_RPMD}
\end{table}%